\documentclass[12pt, english]{article}

\usepackage[english]{babel}  % solve the error: You haven't loaded the option english yet
\usepackage[letterpaper, left=1.1in, right=1.1in, top=1.5in, bottom=1.5in]{geometry}
\usepackage{bm}\usepackage{bbm}   % bm for bold math, bbm for \mathbbm{1}
\usepackage{amsmath}   % for \overset, \underset, \text, \lVert, \rVert
\usepackage{amssymb}   % amssymb or amsfonts for \mathbb
\usepackage{graphicx}  % \includegraphics, \resizebox. graphicx is better than graphics.
\usepackage{cases}     % \begin{cases} for stack of equations after the bracket
\usepackage{multirow}  % \multirow and \multicolumn
\usepackage{longtable} % \begin{longtable} for long tables.
\usepackage{caption}   % \caption*{}, \captionof (while using \resize)
\usepackage{float}     % prevent table from floating by \begin{table}[H]. Using [h] without this package does not avoid the floating.p
\usepackage{natbib}
\usepackage{adjustbox} % \begin{adjustbox} to resize the table
\usepackage[title]{appendix} % more control of appendix
\usepackage{array}     % vertical center in the table \begin{tabular}{ | m{1in} |... }
\usepackage{xcolor}    % change text color
\usepackage{setspace}   %Allows double spacing with the \doublespacing command
\usepackage{url}       % include url: \url{}
\usepackage{pdfpages}    % include pdf pages

\newcommand{\vone}{\mathbbm{1}}      % command to write vector 1 in vector form
\newcommand{\norm}[1]{\left\| #1 \right\|}   % command to write norm
\newcommand{\w}{\widetilde}          % command to write wide tilde

\usepackage{enumitem}  % Change label: \begin{enumerate}[label=(\arabic*)] . Not compatible with lyx: Remove [label=(\arabic*)].
\usepackage{hhline}    % \hhline double lines in tables. Not compatible with lyx: Replace \\hhline\{.*\}\} with \\hline, replace \\hhline\{.*\} with \\hline.

\usepackage{csquotes}     % Overleaf suggest csquotes with biblatex, seems unnecessary.

\usepackage[colorlinks = true,
            linkcolor = blue,
            urlcolor  = blue,
            citecolor = blue,
            anchorcolor = blue]{hyperref}

\begin{document}

\title{\vspace{0cm} The Low-volatility Anomaly and the Adaptive Multi-Factor Model\thanks{We thank Dr. Manny Dong, and all the support from Cornell University. The author names are in alphabetical order. Disclaimer: All opinions and inferences are attributable to the authors and do not represent the views of the T. Rowe Price Associates, Inc.}}
\author{
Robert A. Jarrow\thanks{Ronald P. and Susan E. Lynch Professor of Investment Management, Samuel Curtis Johnson Graduate School of Management, Cornell University, Ithaca, New York 14853, USA and Kamakura Corporation, Honolulu, Hawaii 96815. Email: raj15@cornell.edu.} ,
Rinald Murataj\thanks{Quantitative Analyst at T. Rowe Price, Baltimore, Maryland. Email: rinald.murataj@troweprice.com} ,
Martin T. Wells\thanks{Charles A. Alexander Professor of Statistical Sciences, Department
of Statistics and Data Science, Cornell University, Ithaca, New York 14853, USA. Email: mtw1@cornell.edu.} ,
Liao Zhu\thanks{Ph.D. in the Department of Statistics and Data Science, Cornell University, Ithaca, New York 14853, USA. Email: lz384@cornell.edu.}}

\date{}
\maketitle
\global\long\def\thefootnote{\arabic{footnote}}
%\blindtext

%\tableofcontents
\begin{abstract}
The paper provides a new explanation of the low-volatility anomaly. We use the Adaptive Multi-Factor (AMF) model estimated by the Groupwise Interpretable Basis Selection (GIBS) algorithm to find those basis assets significantly related to low and high volatility portfolios. These two portfolios load on very different factors, indicating that volatility is not an independent risk, but that it's related to existing risk factors. The out-performance of the low-volatility portfolio is due to the (equilibrium) performance of these loaded risk factors. The AMF model outperforms the Fama-French 5-factor model both in-sample and out-of-sample.
\end{abstract}
\textbf{Keywords}: Low-volatility anomaly, AMF model, GIBS algorithm, high-dimensional statistics, machine learning, False Discovery Rate. \\\\
\textbf{JEL}: C10 (Econometric and Statistical Methods and Methodology: General), G10 (General Financial Markets: General)

% \includepdf[pages={1}]{myfile.pdf}
%\includepdf{Disclosure_Bob.pdf}
%\includepdf{Disclosure_Rinald.pdf}
%\includepdf{Disclosure_Marty.pdf}
%\includepdf{Disclosure_Liao.pdf}

\clearpage
\section{Introduction}
The low-risk anomaly is the empirical observation that stocks with lower risk yield higher returns than stocks with higher risk. Here risk is quantified as either a security's return volatility or a security's beta as derived from a Capital Asset Pricing Model (CAPM). This paper focuses only on the
low-volatility anomaly.\footnote{In analyzing the low-beta anomaly, an issue is the non synchronized
trading of small stocks, where a small stock may trade less frequently than the index it is regressed on (market return) to obtain the stock's beta (see \cite{mcinish1986adjusting}).} The low-risk anomaly contradicts accepted APT or CAPM theories that higher risk portfolios earn higher returns. The low-risk anomaly is not a recent empirical finding but an observation documented by a large body of literature dating back to the 1970s. A brief review of the low-risk anomaly literature can be found in appendix \ref{ap: low_risk_review}. Despite its longevity, the academic community differs over the causes of the anomaly. The two main explanations are: 1) it is due to leverage constraints that retail, pension, and mutual fund investors face which limits their ability to generate higher returns by owning lower risk stocks, and 2) it is due to behavioral biases including the lottery demand for high beta stocks, institutions beating index benchmarks with limits to arbitrage, and sell-side analysts over-bias on high volatility stock earnings.

This paper provides a new explanation for the low-volatility anomaly using the newly developed high-dimensional methods based Groupwise Interpretable Basis Selection (GIBS) algorithm with the Adaptive Multi-Factor (AMF) model as \cite{zhu2020high}.

The new insights are due to the advantages of the GIBS algorithm when estimating the AMF model in contrast to the existing methods. The existing methods trace back to the original papers by \cite{ross1976arbitrage} Arbitrage Pricing Theory (APT) and \cite{merton1973intertemporal} Inter-temporal Capital Asset Pricing Model (ICAPM), updated with the recent insights of \cite{feng2017taming}, \cite{kozak2018interpreting}, \cite{giglio2019asset}, and \cite{giglio2020thousands}. These recent papers typically either add factors one-by-one using some refined significance testing procedures, or they use Principal Component Analysis (PCA) or variance-decomposition methods to find a small number of factors. The PCA either mixes the underlying risk-factors or separates them into several principal components, both of which cause interpretation problems. The sparse PCA method used in \cite{kozak2018interpreting} removes some of the weaknesses of the  traditional PCA, and even gives an interpretation of the risk-factors as stochastic discount risk-factors. However, all these methods still suffer from the problems (low interpretability, low prediction accuracy, etc.) inherited from the variance decomposition framework.

The GIBS algorithm changed this framework in the following ways. First, it abandons the variance decomposition framework and is not based on PCA. Instead, it is based on the prototype clustering and high-dimensional methods. It selects basis assets as the ``prototypes'' or ``center'' of the groups they are representing, which gives a clearer interpretation and improved prediction accuracy. Second, instead of searching for a few common risk-factors that affect the entire cross-section of expected returns as in the conventional approach, the GIBS algorithm also finds groupwise basis assets which may only affect stocks in a specific group, sector, or industry. These advantages provide new insights into the low-volatility anomaly.

The GIBS algorithm was developed to estimate the Adaptive Multi-Factor (AMF) model (see \cite{zhu2020high}) which includes both the APT and ICAPM as special cases. The AMF is derived under
a weaker set of assumptions using the recently developed Generalized Arbitrage Pricing Theory (\cite{jarrow2016positive}). Its time-invariance is tested in \cite{zhu2021time}. Its three main benefits are: 1) it is consistent with a large number of risk factors being needed to explain all security returns, 2) yet, the set of risk factors is small for
any single security and different for different securities, and 3) the risk factors are traded. These benefits imply that the underlying model estimated is more robust than those used in the existing literature. In \cite{zhu2020high}, basis assets (formed from the collection of Exchange Traded Funds
(ETF)) are used to capture risk factors in \textit{realized} returns across securities. Since the collection of basis assets is large and highly correlated, high-dimension methods (including the LASSO and prototype clustering) are used. This paper employs the same methodology to investigate the low-volatility anomaly. We find that high-volatility and low-volatility portfolios load on different basis assets, which indicates that volatility is not an independent risk. The out-performance of the low-volatility portfolio is due to the (equilibrium) performance of these loaded risk factors. For completeness, we compare the AMF model with the traditional Fama-French 5-factor (FF5) model, documenting the superior performance of the AMF model.

An outline for this paper is as follows. Section \ref{sec: vol_est_method} discusses the estimation
methodology, and section \ref{sec: vol_estimation_results} presents the results. Section \ref{sec: vol_conclude} concludes. A brief review of the high dimensional statistical methods used in this paper is in Appendix \ref{ap: high_dim_stats}.

\section{The Estimation Methodology}

\label{sec: vol_est_method}

This section gives the estimation methodology. We first pick the universe
of stocks and ETFs, so that we only focus on the assets that are easy
to trade. Then, we form the high and low volatility portfolios and
we choose a time period that exhibits the low-volatility anomaly.
We will employ the Adaptive Multi-Factor (AMF) asset pricing model
with the Groupwise Interpretable Basis Selection (GIBS) algorithm
to these two portfolios and analyze the results in a subsequent section.

\subsection{The Stock and ETF Universe}
\label{sec: vol_universe}

\begin{figure}[H]
\centering
\includegraphics[width=0.7\textwidth, height=3.5in]{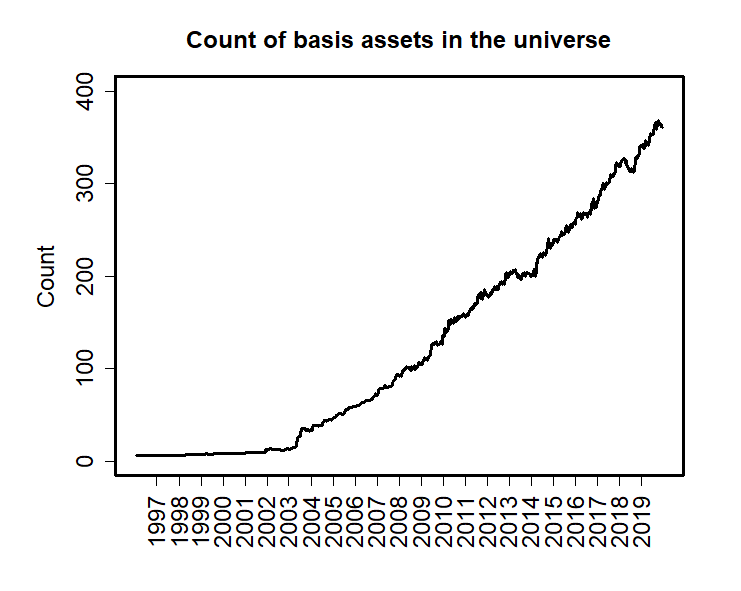}
\caption{Count of the ETFs in the universe.}
\label{fig: vol_etf_count}
\end{figure}

\begin{figure}[H]
\centering
\includegraphics[width=0.7\textwidth, height=3.5in]{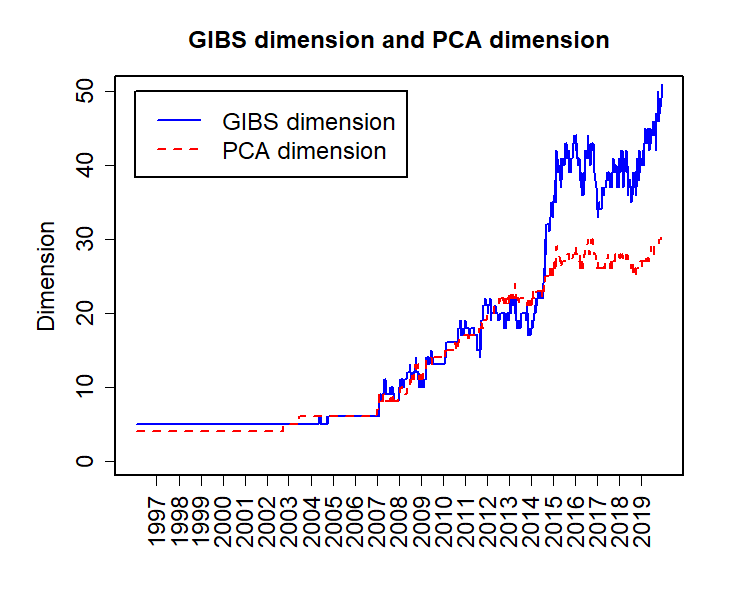}
\caption{GIBS dimension and PCA dimension of ETFs in the universe.}
\label{fig: vol_etf_dim}
\end{figure}

The data initially consists of all stocks and ETFs available in the Center for Research in Security Prices (CRSP) database. To ensure all our securities are actively traded, we focus on stocks and ETFs with a market capitalization ranking in top 2500. This filter excludes small stocks that are more
likely to exhibit the low-volatility anomaly. In addition, we select stocks and ETFs satisfying the following criteria.
\begin{enumerate}
\item According to the description of the CRSP database, ETFs should be with a Share Code (SHRCD) 73. We follow this common practice.
\item We excluded American Depositary Receipts (ADR) from our stock universe. Removing the ADR is a common practice in all empirical finance papers and the main reason is that they are not part of the indices.  This is achieved by using the Share Code (SHRCD) 10 or 11 from the CRSP dataset.
\item We only choose ETFs and stocks which are listed on the NYSE, AMEX and NASDAQ exchanges. This is obtained with the Exchange Code (EXCHCD) 1, 2 or 3 from the CRSP dataset.
\item For a stock to be included at time t, its return has to be observable at least 80\% of trading times in the previous year in order to calculate its volatility. For an ETF to be considered at time t, its return has to be observable during the 3-year regression window before time t.
\end{enumerate}
The number of ETFs in our universe increased rapidly after 2003, see
Figure \ref{fig: vol_etf_count}. To apply the AMF model and GIBS algorithm,
we need enough ETFs to form our collection of basis assets. Hence,
we begin our analysis in 2003. To understand the dimension of the
set of basis assets, we calculate both the GIBS dimension and the
Principal Component Analysis (PCA) dimension of the ETFs in our universe.
The PCA dimension at time $t$ is defined to be the number of principal
components needed to explain 90\% of the variance during the previous
3-years. The GIBS dimension is defined to be the number of ``representatives''
selected using the GIBS algorithm from the basis assets, i.e. the
cardinality of the set $U$ in Table \ref{tab: vol_gibs_algo}. The GIBS
and PCA dimensions of the ETFs across time are shown in Figure \ref{fig: vol_etf_dim}.

Comparing Figures \ref{fig: vol_etf_count} and \ref{fig: vol_etf_dim}, we
see that the number of ETFs and dimensions increase over this time
period. The GIBS and PCA dimension do not increase as fast as the
number of ETFs. But, the GIBS dimension increases faster than the PCA
dimension, suggesting that GIBS is able to pick more basis assets
than does the PCA. The reason is that PCA mixes basis assets together
in linear combinations, while the GIBS algorithm does not.

\subsection{Portfolios and the Low-volatility Anomaly}

\label{sec: vol_anomaly}

To study the low-volatility anomaly, we need to form the high- and low-volatility portfolios. To do this, we first calculate the volatility of the stocks in our universe at time $t$ as the standard deviation of their excess returns over the previous year. The excess return is the raw return minus the risk free rate. Using the excess return is a common practice in the empirical finance literature since it removes the risk free rate and focuses on the risk premiums. The high-volatility portfolio is constructed as an equal-weighted portfolio using the stocks with the highest 25\% volatilities. Similarly, we take the stocks having the lowest 25\% volatilities to form the low-volatility portfolio. To avoid a survivorship bias, we include the delisted returns (see \cite{shumway1997delisting} for more explanation).

\begin{figure}[H]
\centering
\includegraphics[width=0.8\textwidth, height=3.5in]{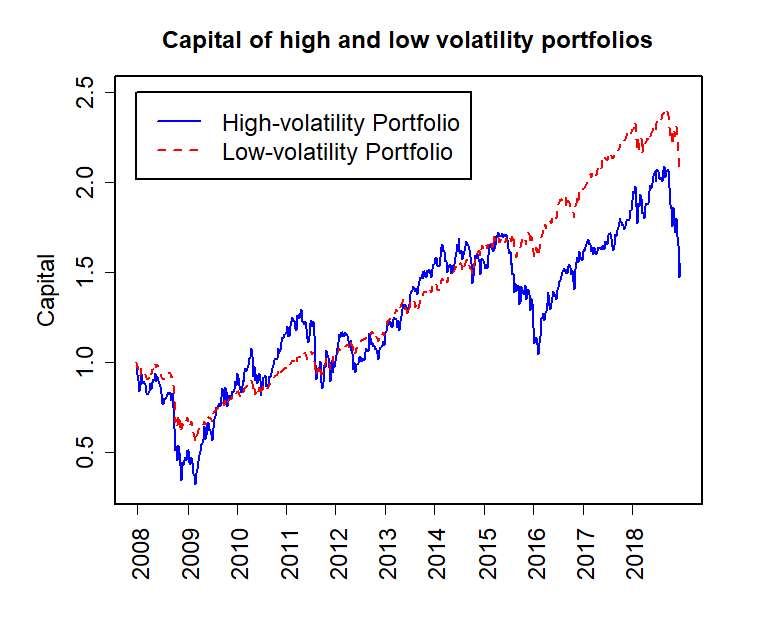}
\caption{Cumulative value of the excess returns from the high- and low-volatility
portfolios.}
\label{fig: vol_caps_ex}
\end{figure}

We then compare the excess returns of the high- and low-volatility
portfolios to verify the existence of the low-volatility anomaly over
our sample period. Given the graph, we selected 2008 as the start
of the anomaly. Between 2008 - 2018, the low-volatility portfolio
had an excess return of 121.4\%, which is higher than the 62.5\% excess
return of the high-volatility portfolio. This documents the existence
of the low-volatility anomaly over our sample period. The superior
performance of the low-volatility portfolio is manifested in Figure
\ref{fig: vol_caps_ex} which graphs the cumulative value of the two portfolios
starting from \$1 at the beginning of 2008.

\subsection{The AMF and GIBS Estimation}
\label{sec: vol_amf_analysis}

This section estimates the Adaptive Multi-Factor (AMF) model based
on the Groupwise Interpretable Basis Selection (GIBS) algorithm proposed
by \cite{zhu2020high}. Although a brief review
of the AMF model and the GIBS algorithm are included in this section,
the details can be found in \cite{zhu2020high}. A sketch of the GIBS
algorithm is in Table \ref{tab: vol_gibs_algo} at the end of this section.

To avoid the effects of market micro-structure frictions, we use a
weekly horizon. Because the number of ETFs increase over time and
the market structure changes during the financial crisis, to approximate
stationarity, we pick a 3-year regression window to do the analysis.
In addition, because the number of ETFs exceed the number of observations
within each regression window, we are in the high-dimensional regime.
Therefore, the high-dimensional GIBS algorithm needs to be used to
estimate the AMF model. We use a dynamic version of the GIBS algorithm
applied to rolling windows. For each week $t$ in 2008 - 2018, we
use the time period from 3 years (156 weeks) window earlier as our current regression
window. We use all the ETFs in our universe as described in Section
\ref{sec: vol_universe} and the FF5 factors as our basis assets. Then,
we apply the GIBS algorithm to select the GIBS determined basis assets
and use them to explain the excess returns of the high- and low-volatility
portfolios using the AMF model. The following is a more detailed introduction
to the AMF model and the GIBS algorithm within each time window.

In the asset pricing theory, given a frictionless, competitive, and
arbitrage free market, a dynamic generalization of \cite{ross1976arbitrage}
APT and \cite{merton1973intertemporal} ICAPM is contained
in \cite{jarrow2016positive}. This extension
implies that the Adaptive Multi-Factor (AMF) model holds for any security's
return over $[t,t+1]$:
\begin{equation}
R_{i}(t)-r_{0}(t)=\sum_{j=1}^{p}\beta_{i,j}\big[r_{j}(t)-r_{0}(t)\big]
=\bm{\beta}_{i}'\cdot[\bm{r}(t)-r_{0}(t)\vone]
\label{eq: vol_linear model}
\end{equation}
where $R_{i}(t)$ denotes the return of the i-th security for $1\leq i\leq N$
(where $N$ is the number of securities), $r_{j}(t)$ denotes the
return on the j-th basis asset for $1\leq j\leq p$, $r_{0}(t)$ is
the risk free rate, $\bm{r}(t)=(r_{1}(t),r_{2}(t),...,r_{p}(t))'$
denotes the vector of security returns, $\vone$ is a column vector
with every element equal to one, and $\bm{\beta}_{i}=(\beta_{i,1},\beta_{i,2},...,\beta_{i,p})'$.

In this paper we are only concerned with the low- and high-volatility
portfolios. Let $R_{1}$ denote the raw return of the low-volatility
portfolio and $R_{2}$ the raw return of the high-volatility portfolio.
To empirically test our model, both an intercept $\alpha_{i}$ and
a noise term $\epsilon_{i}(t)$ are added to expression (\ref{eq: vol_linear model}),
i.e.
\begin{equation}
R_{i}(t)-r_{0}(t)=\alpha_{i}+\sum_{j=1}^{p}\beta_{i,j}(t)\big[r_{j}(t)-r_{0}(t)\big]+\epsilon_{i}(t)
=\alpha+\bm{\beta}_{i}'[\bm{r}(t)-r_{0}(t)\vone]+\epsilon_{i}(t)
\label{eq: vol_testable}
\end{equation}
where $\epsilon_{i}(t)\overset{iid}{\sim}N(0,\sigma_{i}^{2})$ and
$1\leq i\leq N$. The intercept enables the testing for mispriced
securities, and the error term allows for noise in the return observations.

As mentioned earlier, using weekly returns over a 3-year time period
necessitates the use of high-dimensional statistics. To understand
why, consider the following. For a given time period $(t,T)$, letting
$n=T-t+1$, we can rewrite expression (\ref{eq: vol_testable}) as
\begin{equation}
\bm{R}_{i}-\bm{r}_{0}
=\alpha_{i}\vone_{n}+(\bm{r}-\bm{r}_{0}\vone'_{p})\bm{\beta}_{i}+\bm{\epsilon}_{i}
\label{eq: vol_vec}
\end{equation}
where $1\leq i\leq N$, $\bm{\epsilon}_{i}\sim N(0,\sigma_{i}^{2}\bm{I}_{n})$
and {\small{}
\[
\bm{R}_{i}=\begin{pmatrix}R_{i}(t)\\
R_{i}(t+1)\\
\vdots\\
R_{i}(T)
\end{pmatrix}_{n\times1},\,\bm{r}_{0}=\begin{pmatrix}r_{0}(t)\\
r_{0}(t+1)\\
\vdots\\
r_{0}(T)
\end{pmatrix}_{n\times1},\,\bm{\epsilon}_{i}=\begin{pmatrix}\epsilon_{i}(t)\\
\epsilon_{i}(t+1)\\
\vdots\\
\epsilon_{i}(T)
\end{pmatrix}_{n\times1}
\]
}{\small\par}

{\small{}
\[
\bm{\beta}_{i}=\begin{pmatrix}\beta_{i,1}\\
\beta_{i,2}\\
\vdots\\
\beta_{i,p}
\end{pmatrix}_{p\times1},\,\bm{r}_{i}=\begin{pmatrix}r_{i}(t)\\
r_{i}(t+1)\\
\vdots\\
r_{i}(T)
\end{pmatrix}_{n\times1},\,\bm{r}(t)=\begin{pmatrix}r_{1}(t)\\
r_{2}(t)\\
\vdots\\
r_{p}(t)
\end{pmatrix}_{p\times1}\,
\]
}{\small\par}

{\small{}
\begin{equation}
\bm{r}_{n\times p}=(\bm{r}_{1},\bm{r}_{2},...,\bm{r}_{p})_{n\times p}=\begin{pmatrix}\bm{r}(t)'\\
\bm{r}(t+1)'\\
\vdots\\
\bm{r}(T)'
\end{pmatrix}_{n\times p},\,\bm{R}=(\bm{R}_{1},\bm{R}_{2},...,\bm{R}_{N})\label{eq: vol_vec_end}
\end{equation}
}Recall that the coefficients $\beta_{ij}$ are assumed to be constants.
This assumption is only reasonable when the time period $(t,T)$ is
small (say 3 years), so the number of observations is $n\approx150$.
However, the number of basis assets $p$, is around 300 in recent
years, implying that the independent variables exceed the observations.

Because of this high-dimension problem and the high-correlation among
the basis assets, traditional methods fail to give an interpretable
and systematic way to fit the Adaptive Multi-Factor (AMF) model. Therefore,
we employ the Groupwise Interpretable Basis Selection (GIBS) algorithm
to select the basis assets set $S\subseteq\{1,2,...,p\}$ (the derivation
of $S$ is provided later). Then, the model becomes
\begin{equation}
\bm{R}_{i}-\bm{r}_{0}
=\alpha_{i}\vone_{n}+(\bm{r}_{S}-(\bm{r}_{0})_{S}\vone'_{p})(\bm{\beta}_{i})_{S}+\bm{\epsilon}_{i}.
\label{eq: vol_theory_ols_subset}
\end{equation}
The notation $\bm{r}_{S}$ denotes the columns in the matrix $\bm{r}_{n\times p}$
indexed by the index set $S\subseteq\{1,2,...,p\}$, and $(\bm{r}_{0})_{S}$
denotes the elements in the vector $(\bm{r}_{0})_{n\times1}$ indexed
by the index set $S\subseteq\{1,2,...,p\}$. We will use this notation
for any matrices, vectors and indices sets throughout this paper.
An example of expression (\ref{eq: vol_theory_ols_subset}) is the \cite{fama2015five} 5-factor (FF5) model where all of the
basis assets are risk-factors, earning non-zero expected excess returns.
However, FF5 assumes that the number of risk-factors is small and
common to all the securities, whereas the AMF and GIBS does not.

Next, we give a brief review of the GIBS algorithm. For notation simplicity,
denote
\begin{equation}
\bm{Y}_{i}=\bm{R}_{i}-\bm{r}_{0}\,,\quad\bm{X}_{i}=\bm{r}_{i}-\bm{r}_{0}\,,\quad\bm{Y}
=\bm{R}-\bm{r}_{0}\,,\quad\bm{X}=\bm{r}-\bm{r}_{0}
\label{eq: vol_simplify_notation}
\end{equation}
where the definitions of $\bm{R}_{i},\,\bm{R},\,\bm{r}_{i},\,\bm{r}$
are in equations (\ref{eq: vol_vec} - \ref{eq: vol_vec_end}). Let $\bm{r}_{1}$
denote the market return. It is easy to check that most of the ETF
basis assets $\bm{X}_{i}$ are correlated with $\bm{X}_{1}$ (the
market return minus the risk free rate). We note that this pattern
is not true for the other four Fama-French factors. Therefore, we
first orthogonalize every other basis asset to $\bm{X}_{1}$. By orthogonalizing
with respect to the market return, we avoid choosing redundant basis
assets similar to it and also increase the accuracy of fitting. Note
that for the Ordinary Least-Square (OLS) regression, projection does not affect the estimation since it only affects the coefficients, not the estimated $\bm{\hat{y}}$. However,
in high-dimension methods such as LASSO, projection does affect the
set of selected basis assets because it changes the magnitude of shrinking.
Thus, we compute
\begin{equation}
\bm{\widetilde{X}}_{i}=(\bm{I}-P_{\bm{X}_{1}})\bm{X}_{i}
=(\bm{I}-\bm{X}_{1}(\bm{X}'_{1}\bm{X}_{1})^{-1}\bm{X}_{1}')\bm{X}_{i}
\qquad\text{where}\quad2\leq i\leq p_{1}
\label{eq: vol_proj1}
\end{equation}
where $P_{\bm{X}_{1}}$ denotes the projection operator, and $p_{1}$
is the number of columns in $\bm{X}_{n\times p_{1}}$. Denote the
vector
\begin{equation}
\bm{\widetilde{X}}=(\bm{X}_{1},\bm{\widetilde{X}}_{2},\bm{\widetilde{X}}_{3},...,\bm{\widetilde{X}}_{p_{1}}).
\label{eq: vol_proj2}
\end{equation}
Note that this is equivalent to the residuals after regressing other
basis assets on the market return minus the risk free rate.

The transformed ETF basis assets $\bm{\widetilde{X}}$ still contain
highly correlated members. We first divide these basis assets into
categories $A_{1},A_{2},...,A_{k}$ based on a financial characterization.
Note that $A\equiv\cup_{i=1}^{k}A_{i}=\{1,2,...,p_{1}\}$. The list
of categories with more descriptions can be found in Appendix \ref{ap: vol_etf_classes}.
The categories are (1) bond/fixed income, (2) commodity, (3) currency,
(4) diversified portfolio, (5) equity, (6) alternative ETFs, (7) inverse,
(8) leveraged, (9) real estate, and (10) volatility.

Next, from each category we need to choose a set of representatives.
These representatives should span the categories they are from, but
also have low correlation with each other. This can be done by using
the prototype-clustering method with a distance measure defined by
equation (\ref{eq: corr_dist}), which yield the ``prototypes'' (representatives)
within each cluster (intuitively, the prototype is at the center of
each cluster) with low-correlations.

Within each category, we use the prototype clustering methods previously
discussed to find the set of representatives. The number of representatives
in each category can be chosen according to a correlation threshold.
This gives the sets $B_{1},B_{2},...,B_{k}$ with $B_{i}\subset A_{i}$
for $1\leq i\leq k$. Denote $B\equiv\cup_{i=1}^{k}B_{i}$. Although
this reduction procedure guarantees low-correlation between the elements
in each $B_{i}$, it does not guarantee low-correlation across the
elements in the union $B$. So, an additional step is needed, in which prototype clustering on $B$ is used to find a low-correlated representatives
set $U$. Note that $U\subseteq B$. Denote $p_{2}\equiv\#U$.

Recall from the notation definition in equation (\ref{eq: vol_ols_subset})
that $\bm{\w{X}}_{U}$ means the columns of the matrix $\bm{\w{X}}$
indexed by the set $U$. Since basis assets in $\bm{\widetilde{X}}_{U}$
are not highly correlated, a LASSO regression can be applied. By equation
(\ref{eq: lasso_beta}), we have that
\begin{equation}
\bm{\widetilde{\beta}}_{i}
=\underset{\bm{\beta}_{i}\in\mathbb{R}^{p},(\bm{\beta}_{i})_{j}=0 (\forall j\in U^{c})}{\arg\min}
\left\{ \frac{1}{2n}\norm{\bm{Y}_{i}-\bm{\widetilde{X}}_U (\bm{\beta}_{i})_U}_{2}^{2}
+ \lambda\norm{\bm{\beta}_{i}}_{1}
\right\}
\end{equation}
where $U^{c}$ denotes the complement of $U$. However, here we use
a different $\lambda$ as compared to the traditional LASSO. Normally
the $\lambda$ of LASSO is selected by cross-validation. However this
will overfit the data as discussed in the paper
\cite{zhu2020high}. So here we use a modified version of the $\lambda$
selection rule and set
\begin{equation}
\lambda=\max\{\lambda_{1se},min\{\lambda:\#supp(\bm{\widetilde{\beta}}_{i})\leq20\}\}
\end{equation}
where $supp(\bm{\widetilde{\beta}}_{i}) = \{j: (\bm{\widetilde{\beta}_{i}})_{j}\neq0\}$. The notation $\#S$ means the number of elements in the set $S$. Here $\lambda_{1se}$ is the $\lambda$ selected by the ``1se rule''. The ``1se rule'' gives the most regularized model such that error
is within one standard error of the minimum error achieved by the
cross-validation (see \cite{friedman2010regularization,simon2011Regularization,tibshirani2012strong}). Therefore, we can derive the set of basis assets selected as
\begin{equation}
S_i \equiv supp(\bm{\widetilde{\beta}}_{i})
\label{eq: vol_factor_select}
\end{equation}

Next, we fit an Ordinary Least-Square (OLS) regression on the selected basis assets, to estimate $\bm{\hat{\beta}}_{i}$, the OLS estimator from
\begin{equation}
\bm{Y}_{i}=\alpha_{i}\vone_{n}+\bm{X}_{S_{i}}(\bm{\beta}_{i})_{S_{i}}+\bm{\epsilon}_{i}.
\label{eq: vol_ols_subset}
\end{equation}
Note that $supp(\bm{\hat{\beta}}_{i})\subseteq{S}_{i}$. The adjusted
$R^{2}$ is obtained from this estimation. Since we are in the OLS
regime, significance tests can be performed on $\bm{\hat{\beta}}_{i}$.
This yields the significant set of coefficients
\begin{equation}
S_{i}^{*}\equiv\{j:P_{H_{0}}(|\bm{\beta}_{i,j}|\geq|\bm{\hat{\beta}}_{i,j}|)<0.05\}\quad\text{where}\quad H_{0}:\text{True value}\,\,\bm{\beta}_{i,j}=0.
\end{equation}
Note that the significant basis asset set is a subset of the selected
basis asset set. In another words,
\begin{equation}
S_{i}^{*}\subseteq supp(\bm{\hat{\beta}}_{i})\subseteq S_{i}\subseteq\{1,2,...,p\}.
\end{equation}
Then we look at the significant basis assets for the high- and the
low-volatility portfolios separately by creating heatmaps. Each heatmap
presents the percentage of selected factors in all of the ETF sectors.

A sketch of the GIBS algorithm is shown in Table \ref{tab: vol_gibs_algo}.
Recall from the notation definition in equation (\ref{eq: vol_ols_subset})
that for an index set $S\subseteq\{1,2,...,p\}$, $\bm{\w{X}}_{S}$
means the columns of the matrix $\bm{\w{X}}$ indexed by the set $S$.

\begin{table}[H]
\centering
  \begin{tabular}{|| l ||}
  \hhline{|=|}
  \\[-1.6ex]
  \textbf{The Groupwise Interpretable Basis Selection (GIBS) algorithm }  \\[1ex]
  \hhline{|=|}
  \\[-1.6ex]
  Inputs: Stocks to fit $\bm{Y}$ and basis assets $\bm{X}$.
  \\[0.6ex]
  \hline
  \\[-1.6ex]
  1. Derive $\bm{\widetilde{X}}$ using $\bm{X}$ and the Equation (\ref{eq: vol_proj1}, \ref{eq: vol_proj2}).
  \\[0.6ex]
  2. Divide the transformed basis assets $\bm{\widetilde{X}}$ into k groups $A_{1},A_{2},\cdots A_{k}$ using a
  \\[0.6ex]
  \hspace{0.4cm} financial interpretation.
  \\[0.6ex]
  3. Within each group, use prototype clustering to find prototypes $ B_i \subset A_i $.
  \\[0.6ex]
  4. Let $ B = \cup_{i = 1}^k B_i $, use prototype clustering in B to find prototypes $ U \subset B $.
  \\[0.6ex]
  5. For each stock $\bm{Y}_{i}$, use a modified version of LASSO to reduce $\bm{\widetilde{X}}_{U}$
  \\[0.6ex]
  \hspace{0.4cm} to the selected basis assets $\bm{\widetilde{X}}_{S_{i}}$
  \\[0.6ex]
  6. For each stock $ \bm{Y}_i $, fit linear regression on $ \bm{X}_{S_i} $.
  \\[0.6ex]
  \hline
  \\[-1.6ex]
  Outputs: Selected factors $ S_i $, significant factors $ S_i^*$, and coefficients in step 6.
  \\[0.6ex]
  \hhline{|=|}
  \end{tabular}
  \caption{The sketch of Groupwise Interpretable Basis Selection (GIBS) algorithm}
  \label{tab: vol_gibs_algo}
\end{table}

We repeat this estimation process for all of the 3 year rolling regression
windows ending with weeks in 2008 - 2018 (the time period we found
in Section \ref{sec: vol_anomaly} with the low-volatility anomaly). Finally,
we compare the significant basis assets selected for the high- and
low-volatility portfolios. The results are given in the following
Section \ref{sec: vol_estimation_results}.

\section{Estimation Results}
\label{sec: vol_estimation_results}

This section provides the results
from our regressions employing both the FF5 and the AMF model.

\subsection{Residual Analysis: \\Can FF5 explain the low-volatility anomaly?}

We first look at the time series plot of the cumulative capital from
investing in the high- and low-volatility portfolios. Figure \ref{fig: vol_caps_ex}
in Section \ref{sec: vol_anomaly} shows the cumulative capital from the
excess returns for both portfolios from 2008 to 2018. As evidenced
in these graphs, the two portfolios have different volatilities and
the low-volatility portfolio outperforms the high-volatility portfolio.

\begin{figure}[H]
\centering
\includegraphics[width=0.8\textwidth]{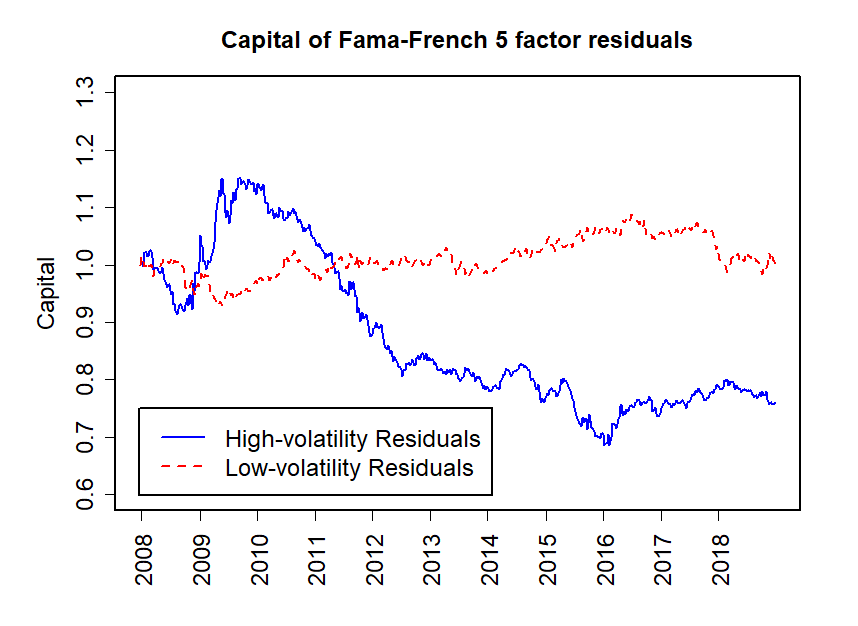}
\caption{Cumulative capital plot of the FF5 residuals of the high-and low-volatility portfolios.}
\label{fig: vol_caps_ff5_resid}
\includegraphics[width=0.8\textwidth]{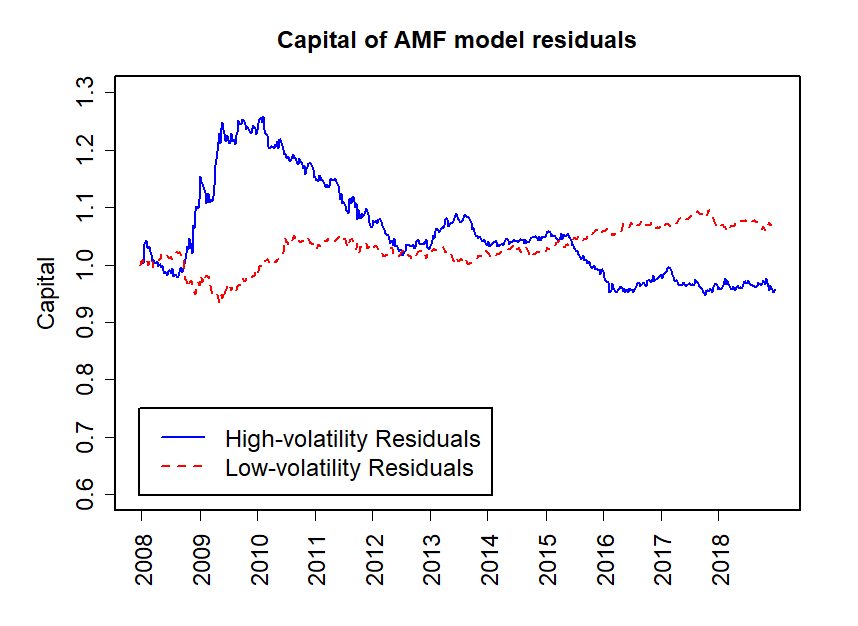}
\caption{Cumulative capital plot of the AMF residuals of the high- and low-volatility portfolios.}
\label{fig: vol_caps_gibs_resid}
\end{figure}

Next, we calculate the cumulative capital from investing in the residual
returns of the high- and the low-volatility portfolios. Figure \ref{fig: vol_caps_ff5_resid}
plots the FF5 model and Figure \ref{fig: vol_caps_gibs_resid} plots the
AMF model. Comparing Figures \ref{fig: vol_caps_ex}, \ref{fig: vol_caps_ff5_resid},
and \ref{fig: vol_caps_gibs_resid}, it is clear that the low-volatility
anomaly is more pronounced in the excess returns as compared to the
residuals. It is still obvious in the FF5 residuals, however, it almost
disappears in the AMF residuals.

Formally, we can test for the differences between the cumulative capitals
from these two portfolios. Since they have different volatilities,
we use Welch's Two-sample t-test corrected for unequal variances.
The hypotheses are
\begin{equation}
H_{0}:\mu_{l}\leq\mu_{h}\qquad H_{A}:\mu_{l}>\mu_{h}\label{eq: vol_resid_test}
\end{equation}
where $\mu_{l}$ indicates the capital of the low-volatility portfolio,
and the $\mu_{h}$ indicates the capital of the high-volatility portfolio.
We do 3 tests where the cumulative capital is calculated with the
excess returns, the residual returns from FF5, and the residual returns
from AMF. If we reject the null-hypothesis $H_{0}$, then there is
strong evidence that the low-volatility anomaly exists.

The p-values of the tests are reported in the parentheses in Table
\ref{tab: vol_resid}. In the Table \ref{tab: vol_resid}, the first row is
the excess return, the second row is the FF5 residual return, and
the third row is the AMF residual return. The first column gives the
return to the low-volatility portfolios from 2008 - 2018. The second
column is that of the high-volatility portfolios. The third column
reports the difference between the two portfolios and gives the p-values
of the tests in equation (\ref{eq: vol_resid_test}).

\begin{table}[H]
\centering
\begin{tabular}{||l|r|r|l||}
\hhline{|*{4}{=|}}
& Low Portfolio & High Portfolio & Low - High (P value) \\
\hhline{|*{4}{=|}}
Excess return & 1.21 & 0.62 & 0.59 ($1.17 \times 10^{-5}$) \\
\hline
FF5 residual return & -0.13 & -0.31 & 0.18 ($9.25 \times 10^{-95}$) \\
\hline
AMF residual return & -0.05 & -0.18 & 0.12 (1.00) \\
\hhline{|*{4}{=|}}
\end{tabular}
\caption{Residual analysis comparing the FF5 and AMF models.}
\label{tab: vol_resid}
\end{table}

The p-value of the FF5 residual is still close to 0, rejecting the
null hypothesis, thereby indicating that the low-volatility anomaly
still exists after adjusting for risk using the FF5 model. In other
words, the FF5 model cannot explain the low-volatility anomaly. However,
the p-value of the AMF residual is close to 1, which implies that
the low-volatility anomaly is not significant after adjusting for
risk with the AMF model. Thus, the AMF model explains the low-volatility
anomaly. Indeed, as we will see in Section \ref{sec: vol_sgn_assets}
below, the AMF model shows that the two portfolios load on different
basis assets (and implied risk factors). It is the (equilibrium) performance
of these factors underlying the low-volatility portfolio that generates
the low-volatility anomaly. Because the FF5 model makes the strong
assumption that every security loads on the same 5 factors, it is
not able to capture the low-volatiity risk premium differences. This
highlights the superior performance of the AMF model.

\subsection{Factor Comparisons}

\label{sec: vol_sgn_assets}

\begin{figure}
\centering
\vspace{-1cm}
\includegraphics[height=10.5cm, width=0.8\textwidth]{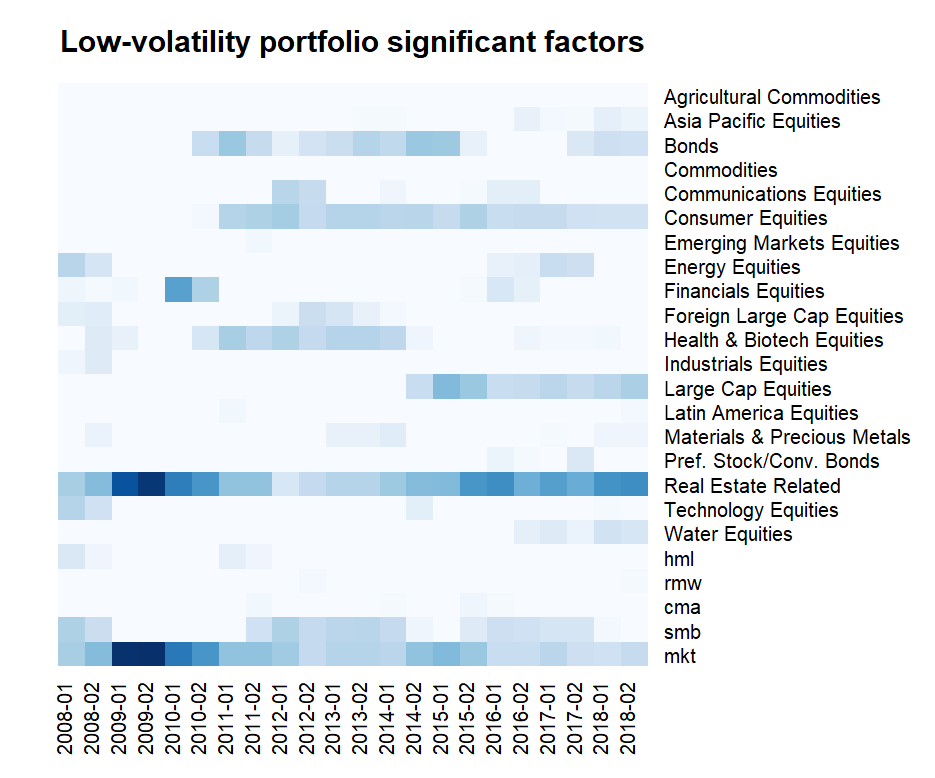} \\
\vspace{-0.5cm}
\caption{Heatmap for the low-volatility portfolio selected factors.}
\label{fig: vol_heatmap_low}
\vspace{0.1cm}
\includegraphics[height=10.5cm, width=0.8\textwidth]{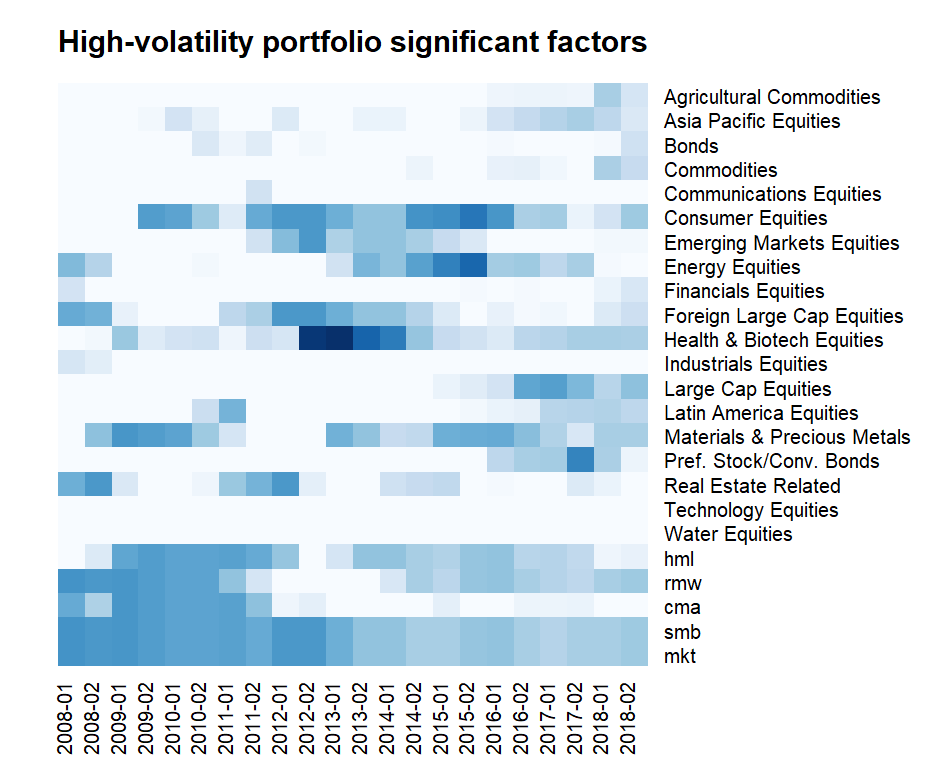} \\
\vspace{-0.5cm}
\caption{Heatmap for the high-volatility portfolio selected factors}
\label{fig: vol_heatmap_high}
\end{figure}

In this section we compare the significant basis assets or ``factors''
selected by the GIBS algorithm for the two high- and low-volatility
portfolios over 2008 - 2018. Figures \ref{fig: vol_heatmap_low} and \ref{fig: vol_heatmap_high}
show the percentage of the significant factors for each ETF class
selected by the GIBS algorithm for the low- and high-volatility portfolios
each half-year across 2008 - 2018 (see Appendix \ref{ap: vol_etf_classes}
for more details on the ETF classifications).

As evidenced in the figures, the two portfolios load on very different
factors. The low-volatility portfolio is mainly related to the ETFs
in Bonds, Consumer Equities, and Real Estate. This is intuitive because
bonds and real estate are of lower risk. Among the FF5 factors, the
low-volatility portfolio only relates to the market return and the
SMB factor. For the high-volatility portfolio, it mainly loads on
ETFs in Materials \& Precious metals, Consumer Equities, Health \&
Biotech Equities, and all the FF5 factors, except CMA. The common
factors for both portfolios are the market return, the SMB factor,
and the Consumer Equity ETFs. Consequently, the out-performance of
low-volatility portfolio is due to the different factor loadings,
and reflects the differential performance of Bonds and Real Estate
relative to materials, precious metals, and the healthcare industry.

Next, we provide a statistical test of whether the two volatility
portfolios load on the same factors. Recalling the notation in equation
(\ref{eq: vol_simplify_notation}), denote $\bm{Y}_{1}$ as the excess return
to the low-volatility portfolio, and denote $\bm{Y}_{2}$ as the excess
return to the high-volatility portfolio over the estimation period.
From equation (\ref{eq: vol_factor_select}), denote $S_{1}$ and $S_{2}$
as the basis assets selected by GIBS for the low- and high-volatility
portfolios, respectively. Then we let
\begin{equation}
\bm{Z}=\begin{pmatrix}\bm{Y}_{1}\\
\bm{Y}_{2}
\end{pmatrix}_{2n\times1},\,\bm{W}=\begin{pmatrix}\bm{X}\\
\bm{X}
\end{pmatrix}_{2n\times1},\,S=S_{1}\cup S_{2}
\end{equation}
\begin{equation}
\bm{W}_{S}=\begin{pmatrix}\bm{X}_{S}\\
\bm{X}_{S}
\end{pmatrix}_{2n\times1},\,\bm{h}=\begin{pmatrix}\bm{0}_{n\times1}\\
\vone_{n\times1}
\end{pmatrix}_{2n\times1}
\end{equation}
The vector $\bm{h}$ is an indicator vector which takes the value
1 for the high-volatility portfolio returns and 0 elsewhere.

The testing for the difference between the basis assets selected by
the two volatility portfolios can be transformed to a testing for
the significance of the interaction between $\bm{h}$ and the selected
factors $\bm{W}_{S}$. We fit two models
\begin{equation}
\text{Model 1: }\bm{Z}=\bm{W}_{S}\bm{\beta}_{S}+\bm{\epsilon}
\end{equation}
\begin{equation}
\text{Model 2: }\bm{Z}=\bm{W}_{S}\bm{\beta}_{S}+\left[\bm{W}_{S}\odot(\bm{h}\vone'_{2n})\right]\bm{\gamma}_{S}+\bm{\epsilon}
\end{equation}
where $\odot$ means the element-wise product for two matrices with
the same dimension. Under the null hypothesis that the two portfolios
have the same coefficients on the same factors, the goodness of fit
of model 1 should be same as that of model 2. So we do an ANOVA(Model1,
Model2) test to compare the two models. The results are in Table \ref{tab: vol_diff_test}.

\begin{table}[H]
\centering
\begin{tabular}{||*{7}{r|}|}
\hhline{|*{7}{=|}}
& Res.Df & RSS & Df difference & Sum of Sq & F statistic & Pr($>$F) \\
\hhline{|*{7}{=|}}
Model 1 & 1139 & 0.24 &  &  &  &  \\
\hline
Model 2 & 1130 & 0.06 & 9 & 0.18 & 355.25 & 0.000 \\
\hhline{|*{7}{=|}}
\end{tabular}
\caption{The ANOVA test of the difference of the factors for the two portfolios.}
\label{tab: vol_diff_test}
\end{table}

The p-value of the test is approximately 0 to 3 decimal places, much
smaller than 0.05, which means that the difference between model 1
and model 2 is significant. This is a strong evidence that the two
volatility portfolios have different factor loadings, which validates
the implications from the heatmap.

\begin{figure}[H]
\centering
\vspace{-0.8cm}
\includegraphics[height=10.5cm, width=0.8\textwidth]{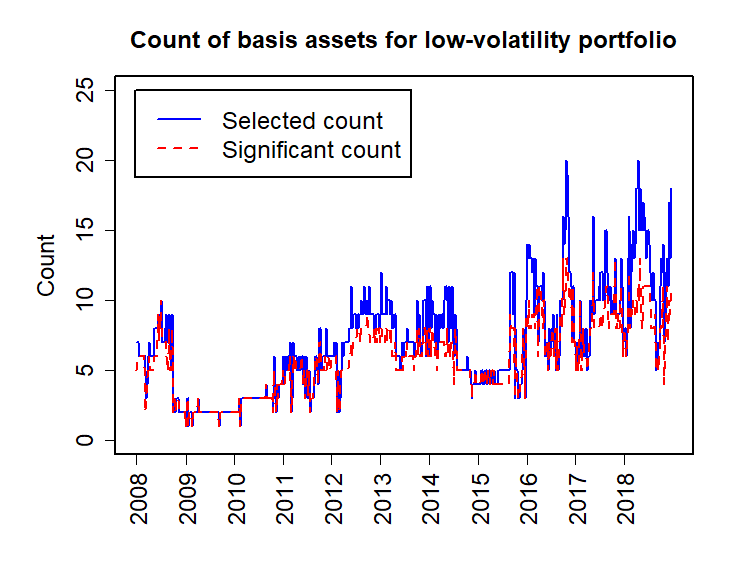} \\
\vspace{-0.8cm}
\caption{The number of selected and significant basis assets for the low-volatility portfolio.}
\label{fig: vol_count_low}
\includegraphics[height=10.5cm, width=0.8\textwidth]{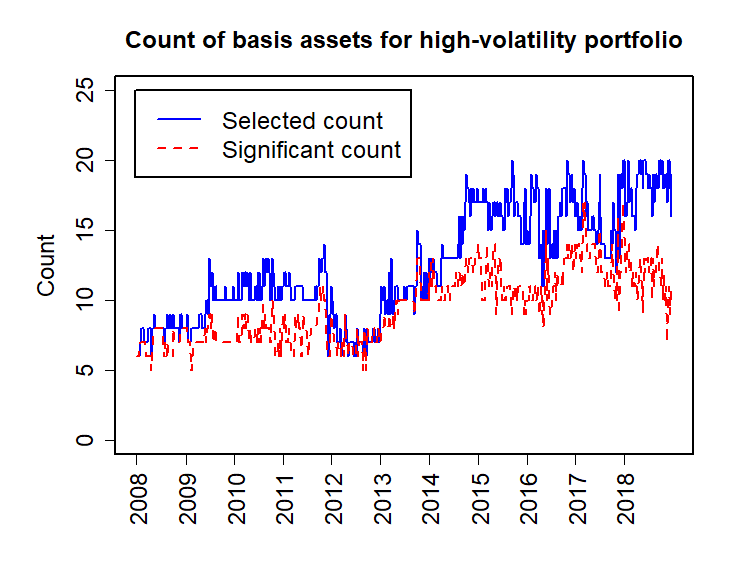} \\
\vspace{-0.8cm}
\caption{The number of the selected and significant basis assets for the high-volatility portfolio.}
\label{fig: vol_count_high}
\end{figure}

Another observation from the heatmaps is that the high-volatility
portfolio is related to more basis assets than is the low-volatility
portfolio. The number of basis assets selected by the GIBS algorithm
and the number of significant basis assets among them is tabulated
in Table \ref{tab: vol_count} and graphed in Figures \ref{fig: vol_count_low}
and \ref{fig: vol_count_high}. We see that more basis assets are selected
and significant for the high-volatility portfolio. This is intuitive
because the larger volatility is generated by the uncorrelated risks
of more basis assets in more volatile industries. More significant
basis assets come from the ETF factors than from the FF5 factors.
On average only 1.54 of the FF5 factors are significant for the low-volatility
portfolio and 3.83 of the FF5 factors are significant for the high-volatility
portfolio, manifesting a limitation of the FF5 model. Furthermore,
most of the ETFs selected by the GIBS algorithm turn out to be significant,
indicating that GIBS is more able to find relevant basis assets.

\begin{table}[ht]
\centering
\begin{tabular}{|| *{7}{r|} |}
\hhline{|*{7}{=|}}
Portfolio & Select & FF5 Select & ETF Select & Signif. & FF5 Signif. & ETF Signif. \\
\hhline{|*{7}{=|}}
Low & 6.96 & 1.72 & 5.23 & 5.82 & 1.54 & 4.29 \\
\hline
High & 12.50 & 4.12 & 8.38 & 9.53 & 3.83 & 5.70 \\
\hline
Difference & 5.54 & 2.39 & 3.15 & 3.71 & 2.29 & 1.41 \\
\hhline{|*{7}{=|}}
\end{tabular}
\caption{The number of the selected or significant basis assets / FF5 factors
/ ETFs for the two volatility portfolios. The ``Select'' column
gives the mean number of basis assets selected by the GIBS algorithm.
The ``Signif.'' column gives the mean number of significant basis
assets among the selected ones. The number of the select / significant
basis assets is the sum of the number of FF5 factors selected / significant
and the ETFs selected / significant. The row ``Low'' is the results
for the low-volatility portfolio, while the row ``High'' is for
high-volatility portfolio. The row ``Difference'' gives the differences
between two portfolios numbers using High - Low.}
\label{tab: vol_count}
\end{table}

In summary, our estimation results show that the two volatility portfolios
load on very different factors, which implies that volatility is not
an independent risk measure, but that it is related to the identified
basis assets (risk factors) in the relevant industries. It is the
(equilibrium) performance of the loaded factors that results in the
low-volatility anomaly and not disequilibrium (abnormal) returns.

\subsection{Intercept Test}

This section provides the tests for a non-zero intercept for both
the AMF and FF5 pmodels for both volatility portfolios. This intercept
test is a test for abonormal performance, i.e. performance inconsistent
with the FF5 and AMF risk models. The null hypothesis is that the
$\alpha$'s in equation (\ref{eq: vol_testable}) are 0.

\begin{figure}
\centering
\vspace{-0.8cm}
\includegraphics[height=10.5cm, width=0.8\textwidth]{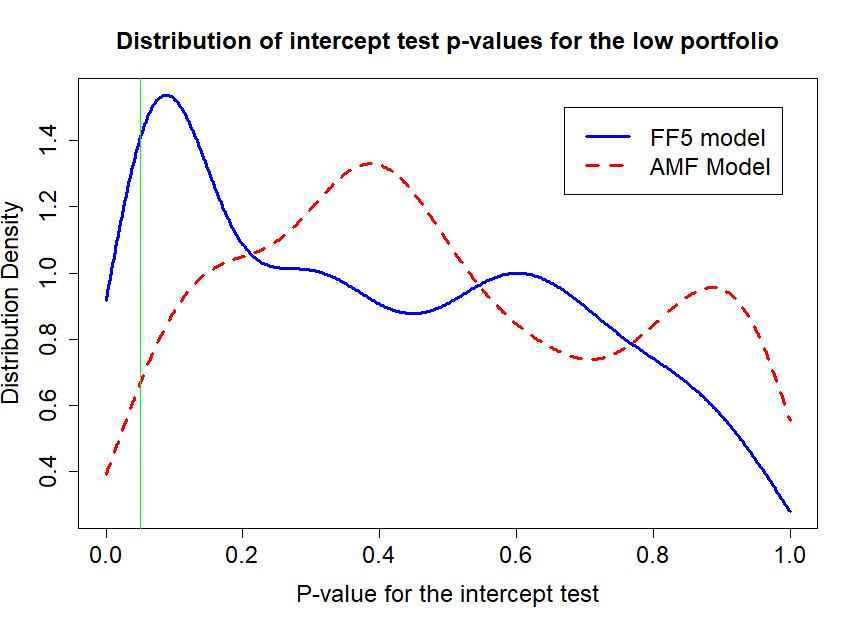} \\
\vspace{-0.6cm}
\caption{Distribution of intercept test p-values for the low-volatility portfolio.}
\label{fig: vol_int_low}
\vspace{0.1cm}
\includegraphics[height=10.5cm, width=0.8\textwidth]{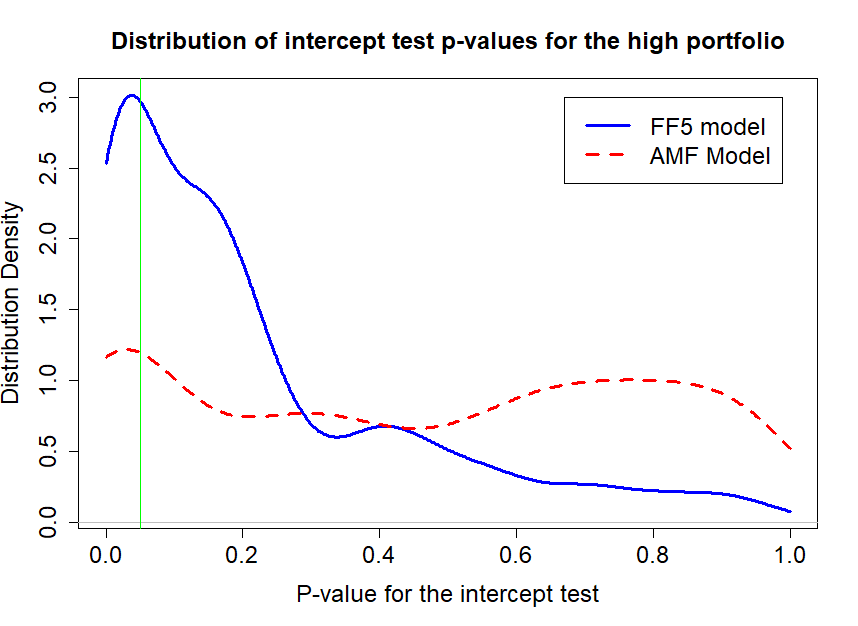} \\
\vspace{-0.6cm}
\caption{Distribution of intercept test p-values for the high-volatility portfolio.}
\label{fig: vol_int_high}
\end{figure}

Figure \ref{fig: vol_int_low} compares the distribution of the intercept
test p-values for the FF5 and AMF models for the low-volatility portfolio,
while Figure \ref{fig: vol_int_high} is for the high-volatility portfolio.
As we can see from the distribution plots, for both portfolios, the
AMF model gives much larger p-values than does the FF5 model. Indeed,
there are more weeks where a significant non-zero intercept is exhibited
with the FF5 model as compared to the AMF model. This suggests that
the AMF model is more consistent with market returns than is the FF5
model and that AMF model does a better job at pricing the anomaly. However, this difference could be due to the fact that we repeat
this test 520 times (520 weeks in 2008 - 2018), and a percentage of
the violations will occur and be observed even if the null hypothesis
is true. Hence, it is important to control for a False Discovery Rate
(FDR).

We adjust for the false discovery rate using the Benjamini-Hochberg-Yekutieli
(BHY) procedures \cite{benjamini2001control} since it accounts for
any correlation between the intercept tests. Note that the Benjamini-Hochberg
(BH) procedure \cite{benjamini1995controlling} does not account for
correlations, and in our case, weekly returns may be correlated. After
adjusting for the false discover rate, for all weeks we fail to reject
the non-zero hypothesis for both the FF5 and AMF models. The results
are in Table \ref{tab: vol_intercept_test}. This evidence is consistent
with the low- and high-volatility portfolios exhibiting no abnormal
performance over our observation period.

\begin{table}
\center %
\begin{tabular}{|| *{5}{r|} |}
\hhline{|*{5}{=}}
\multirow{2}{*}{Portfolio}  & \multicolumn{4}{c||}{Percentage of Significant Weeks}\tabularnewline
\multirow{2}{*}{}  & FF5 p$<$0.05 & AMF p$<$0.05 & FF5 FDR q$<$0.05 & AMF FDR q$<$0.05 \\
\hhline{|*{5}{=|}}
High & 30.1\% & 20.6\% & 0.00\% & 0.00\% \\
\hline
Low & 6.3\% & 2.6\% & 0.00\% & 0.00\% \\
\hhline{|*{5}{=|}}
\end{tabular}
\caption{Intercept Test with control of the False Discovery Rate. The first
row is for the high-volatility portfolio, while the second row is
for the low-volatility portfolio. The columns are related to p-values
and False Discovery Rate (FDR) q-values for the FF5 model and the
AMF model. For each column, we listed the percentage of weeks with
significant non-zero intercept out of all the weeks in the 2008 -
2018 time period.}
\label{tab: vol_intercept_test}
\end{table}

\subsection{In- and Out-of-Sample Goodness-of-fit Tests}

This section compares both the In- and the Out-of-Sample Goodness-of-Fit
tests for both the FF5 and the AMF model. For the In-Sample Goodness-of-Fit
test, for each volatility portfolio, we record the adjusted $R^{2}$'s
(see \cite{theil1961economic}) for both models for each rolling regression.
Then, we calculate the mean of the adjusted $R^{2}$'s. The results
are in the Table \ref{tab: vol_in_sample_gf}.

As shown in this table, even though the FF5 model does a good job
in fitting the data, the AMF model is able to increase the adjusted
$R^{2}$'s. We fit an ANOVA test comparing the FF5 model to the model
using all the basis assets selected by GIBS and FF5. For all the rolling-window
regressions over 2008 - 2018, the p-value of this ANOVA test is close
to 0, less than 0.05. In another words, for all the weeks over 2008
- 2018, the AMF model has a significantly better fit than does the
FF5 model, for both the low- and high-volatility portfolios. Since
we do the tests 520 times (520 weeks in 2008 - 2018), we again need
to adjust for the False Discovery Rate (FDR). However, even after
adjusting the false discovery rate using the most strict BHY method
accounting for the correlation between weeks, the FDR q-value is still
smaller than 0.05 for all of the weeks. This is a strong evidence
that the AMF out-performs the FF5.

\begin{table}
\centering
\begin{adjustbox}{width={0.95\textwidth}, totalheight={0.95\textheight},keepaspectratio}
\begin{tabular}{|| *{5}{r|} |}
\hhline{|*{5}{=|}}
Portfolio & FF5 Adj. $R^2$ & AMF Adj. $R^2$ (change) & $p<0.05$ Ratio & FDR $q<0.05$ Ratio \\
\hhline{|*{5}{=|}}
Low & 0.905 & 0.961 (+6.18\%) & 100\% & 100\% \\
\hline
High & 0.931 & 0.973 (+4.56\%) & 100\% & 100\% \\
\hhline{|*{5}{=|}}
\end{tabular}
\end{adjustbox}
\caption{In-Sample Goodness-of-Fit Tests.}
\label{tab: vol_in_sample_gf}
\end{table}

For the Out-of-Sample Goodness-of-Fit test, we compare the 1-week
ahead Out-of-Sample $R^{2}$ for the FF5 and AMF models for both volatility
portfolios. The results are in Table \ref{tab: vol_out_of_sample_gf}.
As it is seen in the table below, the AMF model out-performs the FF5 model in
the prediction period, which indicates that the better performance
of the AMF model is not due to over-fitting.

\begin{table}
\centering
\begin{tabular}{|| *{3}{r|} |}
\hhline{|*{3}{=|}}
Portfolio & FF5 Out-of-Sample $R^2$ & AMF Out-of-Sample $R^2$ (change) \\
\hhline{|*{3}{=|}}
Low & 0.951 & 0.973 (+2.25\%) \\
\hline
High & 0.973 & 0.982 (+1.01\%) \\
\hhline{|*{3}{=|}}
\end{tabular}
\caption{Out-of-Sample Goodness-of-Fit Tests.}
\label{tab: vol_out_of_sample_gf}
\end{table}

\subsection{Risk Factor Determination}
\label{sec: vol_risk_factor}

The AMF model selects the basis assets that best span a security's
return. Some of these basis assets may not earn risk premium, and
therefore, would not contribute to the risk premium earned by the
volatility portfolios. This section studies which of the ETF basis
assets correspond to risk-factors, i.e. factors that have non-zero
expected excess returns.

At the end of our estimation period 2008 - 2018, there are 335 ETFs
in the universe. Among them, the GIBS algorithm selects 35 ETF representatives
in total. The list of these 35 ETF representatives is contained in
Appendix \ref{ap: etf_representatives} Table \ref{tab: vol_etf_representatives}. Therefore, there are $35+5=40$ basis assets considering both the FF5 factors and ETFs together. In
other words, $p_{1}=335$ and $p_{2}=40$ in Section \ref{sec: vol_amf_analysis}.
For each of these basis assets we compute the sample mean excess return
over our sample period. This is an estimate of the basis assets' risk
premium.

The risk premium estimates for the Fama-French 5 factors are shown
in Table \ref{tab: vol_ff5_risk_premium}. All of these are non-zero.

\begin{table}[H]
\centering
\begin{tabular}{|| *{6}{r|} |}
\hhline{|*{6}{=|}}
Fama-French 5 Factors & Market Return & SMB & HML & RMW & CMAP \\
\hhline{|*{6}{=|}}
Annual Excess Return (\%) & 13.2 & -0.7 & -1.7 & 0.9 & -1.7 \\
\hhline{|*{6}{=|}}
\end{tabular}
\caption{Risk Premium of Fama-French 5 factors}
\label{tab: vol_ff5_risk_premium}
\end{table}

The estimated risk premium for the ETFs are in Table \ref{tab: vol_etf_risk_premium}.
Out of the 35 ETF representatives selected by the GIBS algorithm,
23 of them have absolute risk premium larger than the minimum of that
of the FF5 factors (which is the RMW with absolute risk premium 0.9\%).
This means that at least $(23+5)/(35+5)=70\%$ basis assets are risk
factors in the traditional sense. The list of the 23 ETFs that earns
none-zero risk premium are listed in Table \ref{tab: vol_etf_risk_premium}.

\begin{table}[H]
%\resizebox{6in}{!}{
\centering
\begin{adjustbox}{width={0.95\textwidth}, totalheight={0.95\textheight},keepaspectratio}
\begin{tabular}{||l|l|r||}
\hhline{|=|=|=|}
& & Risk Pre-  \hspace{0.05cm}  \\
ETF name & Category & mium (\%) \\
\hhline{|=|=|=|}
iShares MSCI Brazil ETF & Latin America Equities & 44.0 \\
\hline
VanEck Vectors Russia ETF & Emperging Markets Equities & 20.6 \\
\hline
iShares Mortgage Real Estate ETF & Real Estate & 14.2 \\
\hline
Invesco Water Resources ETF & Water Equities & 13.9 \\
\hline
Vanguard Financials ETF & Financials Equities & 13.2 \\
\hline
Vanguard FTSE Emerging Markets ETF & Emerging Markets Equities & 12.3 \\
\hline
VanEck Vectors Agribusiness ETF & Large Cap Blend Equities & 12.3 \\
\hline
Industrial Select Sector SPDR Fund & Industrials Equities & 12.1 \\
\hline
iShares Select Dividend ETF & Large Cap Value Equities & 11.5 \\
\hline
Materials Select Sector SPDR ETF & Materials & 10.9 \\
\hline
Vanguard Healthcare ETF & Health \& Biotech Equities & 10.2 \\
\hline
iShares U.S. Home Construction ETF & Building \& Construction & 9.5 \\
\hline
iShares MSCI Canada ETF & Foreign Large Cap Equities & 9.4 \\
\hline
SPDR Barclays Capital Convertible Bond ETF & Preferred Stock/Convertible Bonds & 9.2 \\
\hline
First Trust Amex Biotechnology Index & Health \& Biotech Equities & 8.1 \\
\hline
Vanguard FTSE All-World ex-US ETF & Foreign Large Cap Equities & 7.5 \\
\hline
SPDR Barclays High Yield Bond ETF & High Yield Bonds & 6.1 \\
\hline
Invesco DB Agriculture Fund & Agricultural Commodities & -5.9 \\
\hline
iShares MSCI Japan ETF & Japan Equities & 5.8 \\
\hline
iShares MSCI Malaysia ETF & Asia Pacific Equities & 5.7 \\
\hline
Invesco DB Commodity Index Tracking Fund & Commodities & 4.9 \\
\hline
iShares Gold Trust & Precious Metals & 4.2 \\
\hline
Vanguard Consumer Staples ETF & Consumer Staples Equities & 3.8 \\
\hhline{|=|=|=|}
\end{tabular}
\end{adjustbox}
\caption{List of ETFs with large absolute risk premium.}
\label{tab: vol_etf_risk_premium}
\end{table}

\section{Conclusion}
\label{sec: vol_conclude}

In this paper we construct high- and low-volatility portfolios within
the investable universe to explain the low-volatility anomaly using
a new model, the Adaptive Multi-Factor (AMF) model estimated by the
Groupwise Interpretable Basis Selection (GIBS) algorithm proposed
in the paper by \cite{zhu2020high}. For comparison
with the literature, we compare the AMF model with the traditional
Fama-French 5-factor (FF5) model. Our estimation shows the superior
performance of the AMF model over FF5. Indeed, we show that the FF5
cannot explain the low-volatility anomaly while the AMF can. The AMF
results show that the two volatility portfolios load on very different
factors, which indicates that the volatility is not an independent
measure of risk. It is the performance of the underlying risk factors
that results in the low-volatility anomaly. Alternatively stated,
the out-performance of the low-volatility portfolio reflects the equilibrium
compensation for the risk of its underlying risk factors.

%\bibliography{bib_liao}

%\printbibliography

%\bibliographystyle{plain}
%\bibliography{bib_liao}

%\bibliographystyle{ieeetr}
%%\bibliographystyle{chicago}

\bibliography{bib_liao}
\bibliographystyle{chicago}

\begin{appendices}

\clearpage

\section{Low-risk anomaly review}
\label{ap: low_risk_review}

The first explanation of the low-risk anomaly can be traced back to \cite{black1972capitala}
who showed empirically using stock returns from 1926 to 1966 that expected excess returns on high-beta assets are lower than expected excess returns on low-beta stocks (with betas suggested
by the CAPM). In a follow-up paper, accounting for borrowing constraints, \cite{black1972capital} proved that the slope of the line between expected returns and $\beta$ must be smaller than when
there are no borrowing restrictions. More recently, \cite{frazzini2014betting} document the low-beta anomaly in 20 international equity markets and across assets classes including Treasury bonds, corporate bonds, and futures. They argue that investors facing leverage and margin constraints increase the prices of high-beta assets, which generates lower alphas. They show that the Betting Against Beta (BAB) factor, which is long leveraged low-beta assets and short high beta assets, yields positive risk adjusted returns. \cite{ang2006cross} find that stocks with high-idiosyncratic volatility after controlling for size, book-to-market, momentum, liquidity effects, and market-wide volatility risk (VIX) earn lower absolute and risk-adjusted returns than stocks with lower-idiosyncratic volatility.
\cite{ang2009high} also show that there is a strong comovement in the anomaly across 27 developed markets which implies that easily diversifiable factors cannot explain the out performance
of a low idiosyncratic volatility portfolio.

The behavioral explanations (\cite{baker2011benchmarks})
are that irrational investor preferences for lottery-like stocks (more
attention is triggered if you talk about Tesla versus Procter
\& Gamble at a party) and an overconfidence bias pushes the prices
of high risk stocks above their fundamentals. Second, because institutional
investors have a mandate to outperform some market weighted index,
they also over emphasize investments in high-risk stocks. By increasing
the beta exposure of their portfolio in this way, they are more likely
to beat the benchmark in a rising market. And, due to limits to arbitrage,
``smart money'' is not able to arbitrage away this low-risk anomaly.
Providing additional support for this explanation,
\cite{bali2017lottery} show that a proxy for lottery demand stocks
(the average of the five highest daily returns in a given month) explains
this low-beta anomaly. \cite{hong2016speculative}
provide an equilibrium model consistent with this behavior. Last,
it is also argued that the low-risk anomaly is tied to analyst earnings
reports because high-risk stocks are characterized with more inflated
sell side analyst's earnings growth forecasts which produce an investor
overreaction and yields lower returns as shown in
\cite{hsu2013when}.

\section{Prototype Clustering and LASSO}
\label{ap: high_dim_stats}

This section describes the high-dimensional statistical methodologies
used in the GIBS algorithm, including the prototype clustering and
the LASSO. To remove unnecessary independent variables using clustering
methods, we classify them into similar groups and then choose representatives
from each group with small pairwise correlations. First, we define
a distance metric to measure the similarity between points (in our
case, the returns of the independent variables). Here, the distance
metric is related to the correlation of the two points, i.e.
\begin{equation}
d(r_{1},r_{2})=1-|corr(r_{1},r_{2})|\label{eq: corr_dist}
\end{equation}
where $r_{i}=(r_{i,t},r_{i,t+1},...,r_{i,T})'$ is the time series
vector for independent variable $i=1,2$ and $corr(r_{1},r_{2})$
is their correlation. Second, the distance between two clusters needs
to be defined. Once a cluster distance is defined, hierarchical clustering
methods (see \cite{kaufman2009finding}) can be used to organize the
data into trees.

In these trees, each leaf corresponds to one of the original data
points. Agglomerative hierarchical clustering algorithms build trees in a
bottom-up approach, initializing each cluster as a single point, then merging
the two closest clusters at each successive stage. This merging is
repeated until only one cluster remains. Traditionally, the distance
between two clusters is defined as either a complete distance, single
distance, average distance, or centroid distance. However, all of
these approaches suffer from interpretation difficulties and inversions
(which means parent nodes can sometimes have a lower distance than
their children), see \cite{bien2011hierarchical}.
To avoid these difficulties, \cite{bien2011hierarchical}
introduced hierarchical clustering with prototypes via a minimax linkage
measure, defined as follows. For any point $x$ and cluster $C$,
let
\begin{equation}
d_{max}(x,C)=\max_{x'\in C}d(x,x')
\end{equation}
be the distance to the farthest point in $C$ from $x$. Define the
\emph{minimax radius} of the cluster $C$ as
\begin{equation}
r(C)=\min_{x\in C}d_{max}(x,C)
\end{equation}
that is, this measures the distance from the farthest point $x\in C$
which is as close as possible to all the other elements in C. We call
the minimizing point the \emph{prototype} for $C$. Intuitively, it
is the point at the center of this cluster. The \emph{minimax linkage}
between two clusters $G$ and $H$ is then defined as
\begin{equation}
d(G,H)=r(G\cup H).
\end{equation}

Using this approach, we can easily find a good representative for
each cluster, which is the prototype defined above. It is important
to note that minimax linkage trees do not have inversions. Also, in
our application as described below, to guarantee interpretable and
tractability, using a single representative independent variable is
better than using other approaches (for example, principal components
analysis (PCA)) which employ linear combinations of the independent
variables.

The LASSO method was introduced by \cite{tibshirani1996regression}
for model selection when the number of independent variables ($p$)
is larger than the number of sample observations ($n$). The method
is based on the idea that instead of minimizing the squared loss to
derive the Ordinary Least-Square (OLS) solution for a regression, we should add to the loss
a penalty on the absolute value of the coefficients to minimize the
absolute value of the non-zero coefficients selected. To illustrate
the procedure, suppose that we have a linear model
\begin{equation}
\bm{y}=\bm{X\beta}+\bm{\epsilon}\qquad\text{where}\qquad\bm{\epsilon}\sim N(0,\sigma_{\epsilon}^{2}\bm{I}),
\end{equation}
$\bm{X}$ is an $n\times p$ matrix, $\bm{y}$ and $\bm{\epsilon}$
are $n\times1$ vectors, and $\bm{\beta}$ is a $p\times1$ vector.

The LASSO estimator of $\bm{\beta}$ is given by
\begin{equation}
\bm{\hat{\beta}}_{\lambda}=\underset{\bm{\beta}\in\mathbb{R}^{p}}{\arg\min}\left\{ \frac{1}{2n}\norm{\bm{y}-\bm{X\beta}}_{2}^{2}+\lambda\norm{\bm{\beta}}_{1}\right\} \label{eq: lasso_beta}
\end{equation}
where $\lambda>0$ is the tuning parameter, which determines the magnitude
of the penalty on the absolute value of non-zero $\beta$'s. In this
paper, we use the R package \textit{glmnet} \cite{friedman2010regularization}
to fit LASSO.

In the subsequent estimation, we will only use a modified version
of LASSO as a model selection method to find the collection of important
independent variables. After the relevant basis assets are selected,
we use a standard Ordinary Least-Square (OLS) regression on these variables to test for the
goodness of fit and significance of the coefficients. More discussion
of this approach can be found in Zhao, Shojaie, Witten (2017) \cite{zhao2017defense}.

In this paper, we fit the prototype clustering followed by a LASSO
on the prototype basis assets selected. The theoretical justification
for this approach can be found in \cite{reid2018general} and \cite{zhao2017defense}.

\section{Low-correlated ETF representative list}
\label{ap: etf_representatives}

The low-correlated ETF name list in Section \ref{sec: vol_risk_factor} is in Table \ref{tab: vol_etf_representatives}.

{
\small
\begin{longtable}{|| m{0.61\textwidth} | m{0.31\textwidth} ||}
\hhline{|=|=|}
ETF Names  & Category  \\
\hhline{|=|=|}
\endhead
\hline
\multicolumn{2}{c}{{\footnotesize{}Continued on next page}}  \\
\endfoot
\endlastfoot
iShares Gold Trust & Precious Metals \\
\hline
iShares MSCI Malaysia ETF & Asia Pacific Equities \\
\hline
Vanguard FTSE All-World ex-US ETF & Foreign Large Cap Equities \\
\hline
iShares MSCI Canada ETF & Foreign Large Cap Equities \\
\hline
VanEck Vectors Agribusiness ETF & Large Cap Blend Equities \\
\hline
Vanguard FTSE Emerging Markets ETF & Emerging Markets Equities \\
\hline
VanEck Vectors Russia ETF & Emerging Markets Equities \\
\hline
PIMCO Enhanced Short Maturity Strategy Fund & Total Bond Market \\
\hline
iShares 3-7 Year Treasury Bond ETF & Government Bonds \\
\hline
SPDR Barclays 1-3 Month T-Bill ETF & Government Bonds \\
\hline
iShares Short Treasury Bond ETF & Government Bonds \\
\hline
iShares U.S. Home Construction ETF & Building \& Construction \\
\hline
Alerian MLP ETF & MLPs \\
\hline
SPDR Barclays High Yield Bond ETF & High Yield Bonds \\
\hline
Vanguard Healthcare ETF & Health \& Biotech Equities \\
\hline
SPDR Barclays Short Term Municipal Bond & National Munis \\
\hline
Materials Select Sector SPDR ETF & Materials \\
\hline
iShares MSCI Japan ETF & Japan Equities \\
\hline
WisdomTree Japan Hedged Equity Fund & Japan Equities \\
\hline
iShares Mortgage Real Estate ETF & Real Estate \\
\hline
Invesco DB Commodity Index Tracking Fund & Commodities \\
\hline
SPDR S\&P Retail ETF & Consumer Discretionary \hspace{2cm} Equities \\
\hline
Vanguard Financials ETF & Financials Equities \\
\hline
iShares MSCI Brazil ETF & Latin America Equities \\
\hline
iShares MSCI Mexico ETF & Latin America Equities \\
\hline
iShares Select Dividend ETF & Large Cap Value Equities \\
\hline
Invesco Water Resources ETF & Water Equities \\
\hline
SPDR DJ Wilshire Global Real Estate ETF & Global Real Estate \\
\hline
iShares North American Tech-Software ETF & Technology Equities \\
\hline
Consumer Staples Select Sector SPDR Fund & Consumer Staples Equities \\
\hline
SPDR Barclays Capital Convertible Bond ETF & Preferred Stock/Convertible Bonds \\
\hline
Invesco Preferred ETF & Preferred Stock/Convertible Bonds \\
\hline
Invesco DB Agriculture Fund & Agricultural Commodities \\
\hline
Industrial Select Sector SPDR Fund & Industrials Equities \\
\hline
SPDR FTSE International Government Inflation-Protected Bond ETF & Inflation-Protected Bonds\\
\hhline{|=|=|}
\caption{Low-correlated ETF name list in Section \ref{sec: vol_risk_factor}.}
\label{tab: vol_etf_representatives}
\end{longtable}
}

\section{ETF Classes and Subclasses}

\label{ap: vol_etf_classes}

ETFs can be divided into 10 classes, 73 subclasses (categories) in
total, based on their financial explanations. The classify criteria
are found from the ETFdb database: www.etfdb.com. The classes and
subclasses are listed below:
\begin{enumerate}
\item \textbf{Bond/Fixed Income}: California Munis, Corporate Bonds, Emerging
Markets Bonds, Government Bonds, High Yield Bonds, Inflation-Protected
Bonds, International Government Bonds, Money Market, Mortgage Backed
Securities, National Munis, New York Munis, Preferred Stock/Convertible
Bonds, Total Bond Market.

\item \textbf{Commodity}: Agricultural Commodities, Commodities, Metals,
Oil \& Gas, Precious Metals.

\item \textbf{Currency}: Currency.

\item \textbf{Diversified Portfolio}: Diversified Portfolio, Target Retirement
Date.

\item \textbf{Equity}: All Cap Equities, Alternative Energy Equities, Asia
Pacific Equities, Building \& Construction, China Equities, Commodity
Producers Equities, Communications Equities, Consumer Discretionary
Equities, Consumer Staples Equities, Emerging Markets Equities, Energy
Equities, Europe Equities, Financial Equities, Foreign Large Cap Equities,
Foreign Small \& Mid Cap Equities, Global Equities, Health \& Biotech
Equities, Industrials Equities, Japan Equities, Large Cap Blend Equities,
Large Cap Growth Equities, Large Cap Value Equities, Latin America
Equities, MLPs (Master Limited Partnerships), Materials, Mid Cap Blend
Equities, Mid Cap Growth Equities, Mid Cap Value Equities, Small Cap
Blend Equities, Small Cap Growth Equities, Small Cap Value Equities,
Technology Equities, Transportation Equities, Utilities Equities,
Volatility Hedged Equity, Water Equities.

\item \textbf{Alternative ETFs}: Hedge Fund, Long-Short.

\item \textbf{Inverse}: Inverse Bonds, Inverse Commodities, Inverse Equities,
Inverse Volatility.

\item \textbf{Leveraged}: Leveraged Bonds, Leveraged Commodities, \\
 Leveraged Currency, Leveraged Equities, Leveraged Multi-Asset, Leveraged
Real Estate, Leveraged Volati-lity.

\item \textbf{Real Estate}: Global Real Estate, Real Estate.

\item \textbf{Volatility}: Volatility.
\end{enumerate}

In Section \ref{sec: vol_sgn_assets}, we merged several categories to
give a better visualization of the significant factors for each portfolio.
The merged categories are
\begin{itemize}
\item Bonds: Corporate Bonds, Government Bonds, High Yield Bonds, Total
Bond Market, Leveraged Bonds.
\item Consumer Equities: Consumer Discretionary Equities, Consumer Staples
Equities.
\item Real Estate Related: Real Estate, Leveraged Real Estate, Global Real
Estate, Utilities Equities,"Building \& Construction.
\item Energy Equities: Energy Equities, Alternative Energy Equities.
\item Materials \& Precious Metals: Materials, Precious Metals
\item Large Cap Equities: Large Cap Blend Equities, Large Cap Growth Equities,
Large Cap Value Equities.
\end{itemize}

\end{appendices}

\end{document}